# Free space optics in two dimensions: optical elements for silicon photonics without lateral confinement

SIEGFRIED JANZ*, SHURUI WANG, RUBIN MA, JEAN LAPOINTE, AND MARTIN VACHON

*Quantum and Nanotechnology Research Centre, National research Council Canada, Building M-50, 1200 Montreal Road, Ottawa, Ontario, Canada, K1A 0R6*
*siegfried.janz@nrc-cnrc.gc.ca*

**Abstract:** Silicon photonic components based on freely propagating beams in 220 nm thick Si slab waveguides are described and characterized. Examples include optical relays, waveguide crossings, couplers, and resonators with wavelength independent Q factors. The unconfined beams are manipulated using reflecting mirrors etched into the Si layer, which operate in the total internal reflection (TIR) regime. This approach eliminates the back-scattering and waveguide loss arising from sidewall roughness in single mode waveguides, reduces sensitivity to small dimensional variations, and reduces light induced self-heating. Although the detailed behavior of TIR at curved waveguide side wall mirror is too complex to capture by simple models, the experimental results are found to be in qualitative agreement with simple analytical calculations based on Gaussian beam theory and effective index approximations.

## 1. Introduction

This work describes a selection of silicon photonic components based on freely propagating light beams in Si slab waveguides, that function without the use of lateral confinement. Single mode waveguides are the basic connecting element used in almost all Si photonic devices. In the most common configuration, the waveguides are etched into a 220 nm thick Si layer on a silicon-on-insulator (SOI) wafer, and must be less than 500 nm wide so that only a single mode is supported. The small size and high index contrast of Si waveguides have enabled many optical devices with a very small foot-print, but also make it challenging to produce devices repeatably to a given performance specification. Fabricating Si devices with dimensional fidelity to design of better than 10 nm accuracy is usually difficult. Even such small deviations in feature size from the nominal design often have a noticeable effect on device performance due to the resultant change in waveguide effective index $N_{eff}$ and hence propagation phase delay. These phase errors can limit the performance in devices that rely on precise phase relationships between two or more single mode waveguides such as arrayed waveguide grating (AWG) wavelength demultiplexers [1] and Mach-Zehnder interferometers (MZI) [2]. In a silicon waveguide, the typical etched side wall roughness of a few nanometers produces significantly higher waveguide scattering losses than in lower index contrast waveguides based on silicon nitride or silicon dioxide [3,4]. In the case of ring resonators, this scattering not only limits the achievable Q-factor, but also couples light into counterpropagating modes resulting in resonance line shape distortion and line splitting [5-7]. Finally, optical powers less than 1 mW concentrated in a single mode Si waveguide can produce self-heating and other nonlinear optical effects. While such nonlinear effects are of interest for some applications, they can also cause a power dependent distortion of the output signal in Si photonic sensors and wavelength filters where reproducibility and calibration accuracy are essential [1,7].

These issues have motivated work on reducing the sidewall roughness by using surface oxidation smoothing techniques to produce extremely smooth sidewalls [8,9]. While successful in reducing waveguide loss, such fabrication focused strategies add complexity to the fabrication process, and also do not necessarily improve variability in fabricated dimensions. The use of wider multimode waveguides is a different strategy to reduce the interaction of light with the etched Si sidewall. [2,10,11]. These ring resonator and MZI devices clearly demonstrate that by distancing the sidewall from the optical path, side wall scattering is reduced and device performance is significantly improved. Since these devices support several waveguide modes, optical mode control and filtering through device design or active mode manipulation is required.

Another strategy is to employ freely propagating beams in slab waveguides without relying on laterally confined waveguides in any critical functional role. This approach eliminates the scattering loss, phase error and back-reflection due to waveguide sidewall roughness. Any remaining residual scattering from the SOI wafer surface is much smaller, since the surface roughness of manufactured SOI wafers is typically less than a nanometer. Silicon echelle grating waveguide demultiplexers are a well-established example of the planar waveguide approach, in which the dispersive structure is a planar slab waveguide bounded by a shaped reflective diffraction grating [12]. Echelle demultiplexers do not suffer from the phase errors and loss resulting from propagation through the single mode waveguide array in an AWG demultiplexer. Recently, mode converters and beam expanders using parabolic sidewall reflectors to expand light from a single mode Si waveguide have been demonstrated by two groups [13,14]. These structures consist of slab waveguide free propagation region (FPR) bounded by a parabolic sidewall mirror. This component can expand the beam from a single mode waveguide to any size, for example to achieve a better match to an optical fiber or a create a free space beam, thereby offering a more compact and flexible alternative to the tapered waveguides or focusing gratings used in most fiber-chip couplers. The insertion loss of these mode converters has been measured to be as low as 0.15 dB [13]. The parabolic beam expander has much more general applications than just fiber coupling, since it can create a collimated slab mode beam of any desired width from a single mode waveguide input [14] and, by reciprocity, capture a collimated slab beam and couple the light efficiently back into a single mode waveguide. These collimators therefore act as an almost lossless interface between single mode waveguide optics and free propagating slab mode optics. For example, parabolic collimators are were recently used to launch and capture the input and output light in a silicon circulating Gaussian beam resonator (CGBR) based on free propagating slab modes [15]. The CGBR has no internal single mode waveguides or waveguide directional couplers, and exhibited very promising performance compared with conventional Si ring resonators. Quality factors of more than Q=300000 were achieved, and mode splitting due to back-scattering was eliminated.

The results reported in this paper are intended to complement the recent work on the CGBR resonator [15] by providing a more detailed measurement and analysis of the individual elements that were combined in the CGBR resonator, as well as presenting two new functionalities: the optical relay and waveguide crossing. In particular, we present the designs and measured characteristics of optical relays for beam transmission within the Si chip, optical waveguide crossings, and optical beam splitter couplers (BSC). In the final section we employ a similar CGBR resonator to that of Ref. [15] to demonstrate the wavelength independence of a BSC coupler used as a through and drop port in a resonator. These optical elements use of total internal reflection (TIR) mirrors to manipulate and direct light in the slab waveguides. In a 220 nm thick Si waveguide with an effective index of approximately $N_{eff} = 2.8$, an incident slab mode beam will undergo lossless TIR at an etched sidewall provided the incident angle exceeds the critical angle of $\theta_c = 21°$ for an Si-air sidewall boundary (i.e., a Si waveguide with no upper cladding) or $\theta_c = 31°$ for a Si-SiO$_2$ sidewall boundary (i.e., a Si waveguide with an SiO$_2$ upper cladding). Lossless TIR over such a wide range of incident angles enables the practical application straight and curved sidewall mirrors to direct and shape freely propagating slab mode beams in a wide range of configurations. Since TIR is imposed by overall wavevector conservation at the reflecting edge, the details of the etched sidewall vertical profile including slope will not modify the TIR angle range or reflection loss.

## 2. Fabrication and Measurement

The structures characterized in this work were all fabricated on a silicon-on-insulator chip with a 220 nm thick Si waveguide layer on top of a 2 μm thick buried SiO$_2$ oxide layer that forms the lower cladding. Electron beam lithography and reactive ion etching were used to pattern and fabricate the structures. All waveguides, coupling gratings and slab waveguide devices were formed in a single step by etching away the silicon layer through to the buried oxide layer outside the Si feature boundaries. The upper cladding was added by depositing a 2 μm thick layer of SiO$_2$ by plasma enhanced chemical vapor deposition (PECVD).

The measurements were carried out using a fiber-coupled tunable laser with a wavelength tuning range from $\lambda = 1460$ nm to $\lambda = 1580$ nm, and an in-fiber incident power of 1 mW. Transverse electric (TE) light was coupled into and from the on-chip waveguides through single mode fibers terminated with 8° angle polished facets, that were positioned directly above focusing grating couplers fabricated on the chip surface. The output fiber was directed to a fiber-coupled photodetector to measure the transmitted optical power. All measurements were carried out using the TE polarized waveguide modes on chip.

The optical components that were characterized were all connected to the input and output grating couplers by 500 nm wide single mode Si channel waveguides. The focusing grating coupler design has been previously described in Ref. [16] and [17]. The fabricated single mode waveguide loss and grating coupler efficiency were determined by comparing the insertion loss of three waveguides of lengths 6.1 mm, 17 mm and 31 mm. The slope of insertion loss with length provides the waveguide loss, while the zero-length intercept yields the input and output grating coupling efficiency. The waveguide loss of the 500 nm wide single mode connecting waveguides was approximately -1 dB/cm. The grating coupler efficiency was wavelength dependent, with measured peak coupling efficiency of approximately -7 dB near $\lambda=1530$ nm, and a -3 dB wavelength bandpass width of approximately 50 nm. The device measurements presented in this work were all normalized to the transmission of a straight reference waveguide connected to an input and an output grating coupler. A reference waveguide spectrum example is shown in Fig. 2(a) below. This normalization procedure removes the input and output waveguide loss and the wavelength dependent grating coupler loss from the measurement data, so that the results represent the device insertion loss alone. The peak transmission of identical reference waveguides at different locations across the chip was observed to vary within a range of ±1 dB and also shift by a few nanometers in wavelength, due to small fabrication variations across the chip. The normalized data therefore has a corresponding ±1 dB uncertainty. The coupling peak wavelength shift has little effect near the coupling efficiency maximum where the wavelength variation is nearly flat, but can introduce spurious wavelength variations in the normalized data at the short and long wavelength tails of the grating coupling envelope where the coupling efficiency decays steeply.

## 3. Planar Waveguide Relay Optics.

A pair of identical parabolic mirrors facing each other can map the input light distribution at the focal point of the first parabola onto the focal point of the second parabola. Such a relay pair can therefore transform a waveguide input mode to a wide collimated beam, which is then captured and efficiently coupled into an output waveguide a considerable distance away, without the use of long single mode waveguides. The use of directed slab beams may also simplify optical circuit architecture since the need for waveguide crossings is eliminated. Light beams simply pass through each other in a slab waveguide.

Three sets of optical relays were designed and fabricated for free propagation lengths of L=200 μm, L=800 μm and L=3200 μm, as shown in Fig. 1. Each optical relay consists of a pair of identical parabolic mirror beam collimators, denoted here as the transmitter and receiver. The input waveguide aperture is positioned at the transmitter parabola focal point, so that the light emitted from the waveguide diverges and is then reflected from the parabola edge to form a wide collimated beam that is directed towards the receiver parabola through the Si slab waveguide free propagation region (FPR) of length L. The receiver parabola captures the collimated beam and refocuses the light into the output waveguide. These relays are similar in layout to the test structures employed in Ref. [13] to measure beam expander insertion loss, but here the layout is designed to assess the variation in transmission with propagation distance and parabola focal length. For each propagation length, four relays with parabolic collimators with parabola focal lengths F=5 μm, F= 10 μm, F=20 μm, F = 50 μm were fabricated, and arranged on chip as shown in Fig. 1. The device layout in Fig. 1 represents the actual scale and proportions of each device shown, as fabricated on chip.

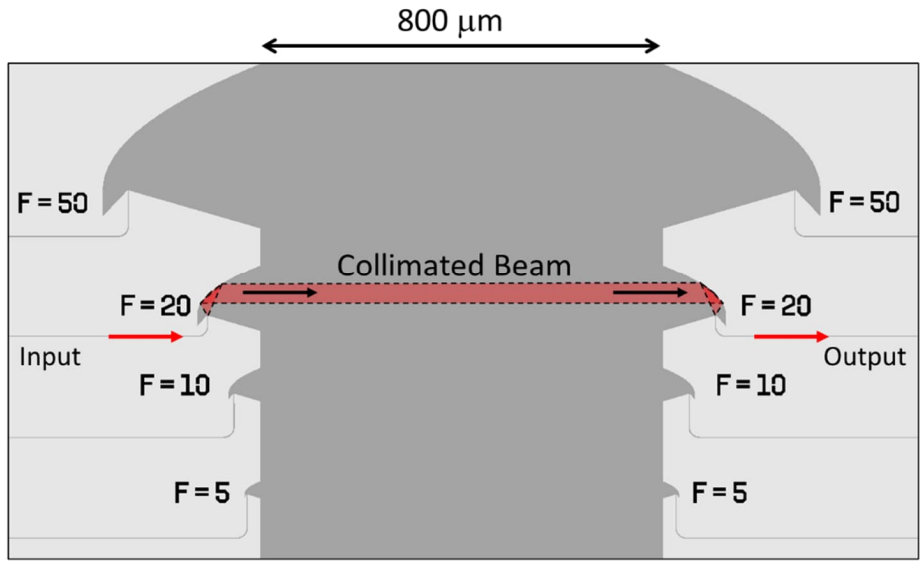

Fig. 1. Layout of the optical relay test structures for an 800 μm long free propagation pathlength. The dark grey areas indicate the 220 nm thick Si waveguide regions. The four separate relay pairs with collimator focal lengths ranging from F = 5 μm to F= 50 μm direct light across common free propagation region. The collimated beam path through the F= 20 μm relay is indicated by the red shaded area outlined by dashed lines.

The performance of the collimator and relay can be qualitatively understood using Gaussian beam theory in two dimensions [18] to describe beam propagation in the slab waveguide. Accurate numerical simulation over large free propagation lengths and Si feature scale variations can be prohibitively time consuming. Although the lateral field profiles of the input waveguide mode and collimated beams do not have exact Gaussian profiles, the profiles are qualitatively similar. At the transmitter collimator input, the light in the single mode 500 nm wide input waveguide is first expanded by a 65 μm long waveguide taper that varies from the initial 500 nm width to a 1 μm aperture at the parabola focal point, where it is launched into transmitter parabolic collimator section. This wider aperture is used to reduce the beam divergence of the light entering the slab waveguide. The input waveguide mode at the 1 μm wide input waveguide aperture can be approximated by a Gaussian profile with beam spot size of $\omega_0 = 0.38$ μm. This is distance from the beam center to the 1/e field amplitude point of the first order TE mode of the 1 μm × 0.22 μm Si waveguide at a wavelength of $\lambda=1550$ nm, as calculated using a commercial finite difference mode solver. The resulting beam divergence half-angle in the FPR is estimated to be 26°, again using the Gaussian beam model. The reflecting parabola edge in direct line with the input waveguide is at a 45° angle to the central incident ray of the diverging beam. Due to the curvature of the parabolic mirror, all the rays within the ±26° angular spread are incident on the parabola edge at angles between 33° and 58°. For a waveguide without cladding, the critical angle of the slab waveguide to air interface is approximately $\theta_c = 21°$, which increases to $\theta_c = 31°$ when a $SiO_2$ cladding is deposited. Therefore, for both clad and unclad Si waveguides, almost all the light is incident on the parabola edge at angles larger than the critical angle and will undergo lossless total internal reflection (TIR).

Transmission spectra from $\lambda = 1480$ nm to $\lambda = 1580$ nm were collected for each relay and normalized by the transmission spectrum of a straight reference waveguide as outlined in Section 2. Fig. 2(a) shows some example transmission spectra for different focal length along with the as measured reference waveguide spectrum that is used for normalization. No obvious wavelength dependence of the relay signal was observed near the grating coupling peak. The upward slope at long and short wavelengths is an artefact of normalization due to variations in input and output grating coupler envelope shape across the chip, as noted in Section 2. In Fig. 2(b), the data points are an average of the relay transmission spectrum between $\lambda = 1510$ and 1540 nm encompassing the range of optimal coupling efficiency of the coupling gratings. For the largest two parabolas, the relay transmission loss in Fig. 2 shows very little change with increasing propagation distances up to 3200 μm. For shorter focal lengths, the transmitted power clearly decreases with propagation distance. For $F = 5$ μm relay, the transmitted power decreases with length and is -10 dB lower at L=3200 μm than for L=200 μm.

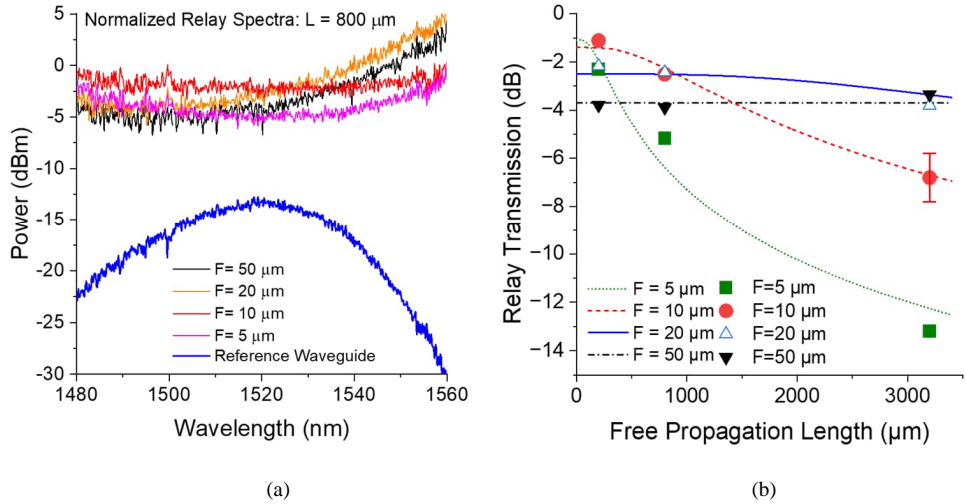

(a)  (b)

Fig. 2. (a) The normalized relay transmission spectra for each collimator focal length and an L= 800 μm free propagation length. The as measured refence waveguide spectrum is also shown. (b) Relay transmission loss variation with free propagation length for each collimator focal length. The curves indicate the transmission loss calculated using the planar Gaussian beam model described in the text. The error bar indicates the estimated uncertainty in each data point due to variations in grating coupler efficiency for different devices.

The variation of relay transmission with focal length can be attributed to the intrinsic divergence of the collimated beams with propagation distance. Since the transmitter and receiver collimator structures are identical, efficient coupling of the relayed beam to the output waveguide requires that the incident lateral beam profile arriving at the receiver parabola closely matches the initial launched beam width and shape at the transmitter parabola. In the following analysis, the mode launched into the collimator at the 1 μm wide input waveguide aperture is again assumed to have a Gaussian beam waist of $\omega_0 = 0.38$ μm, based on a mode solver calculation, as described above. This gives a Gaussian confocal beam parameter of $z_0 = 1.42$ μm, where

$$z_0 = \frac{\pi \omega_0^2 N_{eff}}{\lambda}. \qquad (1)$$

is propagation distance over which the beam spot size $\omega(z)$ increases from $\omega_0$ to $\omega(z) = \sqrt{2}\, \omega_0$. Once $z_0$ is known, the light distribution incident on the transmitter parabola edge and hence the collimated beam half width $\omega(2F)$ can be estimated assuming Gaussian beam propagation [18], where 2F is the distance from the input waveguide aperture to the parabola edge. The subsequent beam spot size variation $\omega(z)$ as the collimated beam propagates through the FPR waveguide towards the receiver parabola can also be modelled using Gaussian beam propagation, but now taking $\omega_0$ as the collimated beam half width at the transmitter parabola. Note that the full beam width $D(z) = 2\cdot\omega(z)$ between opposite 1/e field amplitude points is twice the Gaussian beam spot size $\omega(z)$. The confocal parameter and beam widths $D(z)$ of the launched beam (i.e., at L=0) and the received beam at the transmitter are given in Table 1 for each relay length L and parabola focal length.

Table 1. Calculated relay confocal parameter $z_0$ and full beam widths $D(z)$ for different propagation lengths.

| Focal Length (μm) | $z_0$ (μm) | $D(z)$ z = 0 μm (μm) | $D(z)$ z = 200 μm (μm) | $D(z)$ z = 800 μm (μm) | $D(z)$ z = 3200 μm (μm) |
|---|---|---|---|---|---|
| 5 | 123 | 9 | 18 | 61 | 243 |
| 10 | 489 | 19 | 20 | 36 | 123 |
| 20 | 1953 | 37 | 37 | 40 | 71 |
| 50 | 12203 | 93 | 93 | 93 | 96 |

The coupling efficiency into the output waveguide at the receiver collimator at distance L can be estimated the mode overlap integral of the relayed beam with the lateral mode width of the launched beam, since the two parabolas are identical. A planar lateral Gaussian field profile has the form $E(x) \sim \exp(-x^2/\omega(z)^2)$ where x is the lateral distance from the beam axis and $\omega(z)$ from is the Gaussian beam half-width at a distance z along the propagation direction. The mode overlap integral of two such beam profiles with spot sizes $\omega_1$ and $\omega_2$ has the analytical form

$$\Lambda = 2 \cdot \frac{\omega_1 \omega_2}{\omega_1^2 + \omega_2^2}. \qquad (2)$$

The variation of the calculated Gaussian overlap loss $\Lambda$ with relay length is shown in Fig. 2 for the four different collimator focal lengths. For each focal length, a constant length independent loss was added to the calculated loss to account for experimental losses not related to propagation. This offset was determined by minimizing the square error between the experimental data points and the calculated curve for each focal length. Extrapolating the model curves to the L=0 μm intercept indicates that the background insertion loss of each relay is better is better than -0.5 dB per collimator for the F=5 μm relay, and increases with collimator size to almost -2 dB for the F=50 μm relay. Possible reasons for this loss will be considered in the Section 6. The length dependence of the measured relay transmission and Gaussian model are in reasonable agreement, given the ±1 dB measurement uncertainty.

A Gaussian beam with a beam waist $\omega_0$ at z = 0 will have a spot size $\omega(z_0) = \sqrt{2}\cdot\omega_0$ after propagating a distance $z_0$. Using Eq. 2 the overlap integral will have a very low value of $\Lambda=0.94$ or -0.26 dB. In the Gaussian approximation, the confocal parameter is therefore a useful guide in choosing the collimated beam diameter to give low loss for any specific relay length. For the F=50 μm relays, the confocal parameter $z_0 = 1.2$ cm is much longer that the longest propagation length of 3200 μm, and there is negligible length dependent loss in the calculated or experimental results for all three relay lengths. For the shorter focal length collimators, the loss becomes significant when the relay length exceeds the confocal parameters given in Table 1.

## 4. Beam crossing

Parabola-based relay optics can be used to transmit beams over long distances with negligible loss due to etched sidewall scattering. Another obvious but useful feature is that independent beams can cross paths with no cross-talk, back-reflection, and scattering loss arising from discontinuities at intersecting waveguides. In complex optical circuits, the layout architecture is often constrained by the desire to avoid optical path crossings. Many low loss waveguide crossing designs have been described [19-21]. Through careful design and simulation, low loss and cross talk can be achieved, but are never entirely eliminated. On the other hand, the crossing of two free propagating beams in a planar slab waveguide is intrinsically lossless and introduces no back reflection. If light is transmitted across a chip using relay optics, a specific waveguide crossing structure is obviously unnecessary. However, in situations where single mode waveguides are used and must cross paths, the relay optics structures can be combined to create a simple waveguide crossing structure as shown in Fig. 3.

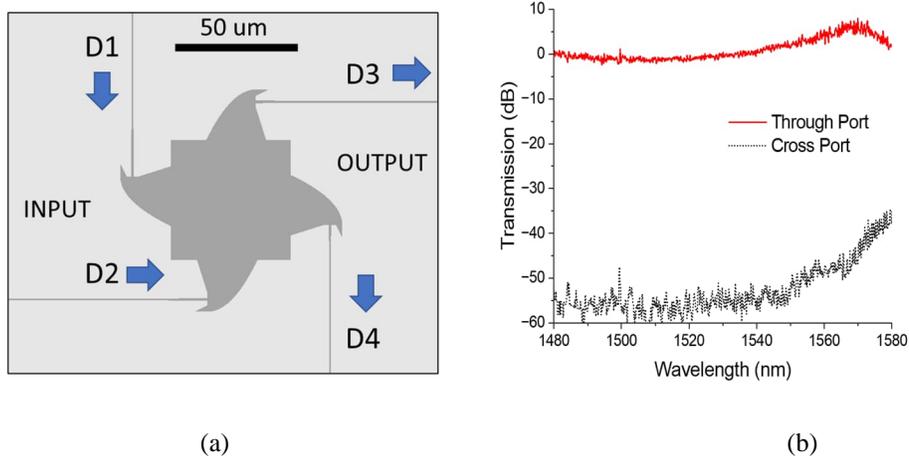

(a)            (b)

Fig. 3. (a) Top view of waveguide crossing test structure layout consisting of four parabolic collimators with a focal length of F=5 μm. The dark grey areas indicate the boundaries of 220 nm thick Si slab waveguide regions. (b) The measured through port (D2-D3) and cross port (D1-D3) transmission for the beam crossing structure.

The primary requirement for lossless crossing is that the parabolic collimator separation is less than the confocal parameter $z_0$ of the collimated beams, as demonstrated in Section 3. The waveguide crossing shown in Fig. 3(a) was fabricated on the same SOI chip with 220 nm Si waveguide as described in Section 2. The crossing consists of four parabolic collimators arranged on the sides of a 50 μm square. Each collimator collimates a beam of light from one input waveguide and directs it to the collimator on the opposite side of the square, where the light is refocused back into the output waveguide. The collimators are identical to the F = 5 μm collimators described in Section 2, with the 500 nm wide input waveguide again tapered out to a 1 μm aperture at the parabola focal point. The parabola confocal parameter, $z_0$ =123 μm, is twice the crossing distance, so that losses due to beam divergence should be negligible.

Crossing measurements were carried out over a wavelength range from λ = 1460 nm to λ =1580 nm and normalized to a straight reference waveguide spectrum again using the methodology as outlined in Section 2. Light was coupled into each input waveguide D1 and D2, and the received light on the output side was measured in outputs D3 and D4. The D2-D3 spectrum gives the through port insertion loss, while the D1-D3 spectrum gives the cross-port rejection. The through port and cross-port spectra are shown in Fig. 3(b). The through port insertion loss from λ =1500 to λ =1550 nm is within ±1 dB of zero, while the measured cross port rejection ratio near -55 dB is limited by the noise floor of the measurement system. At the short wavelength and long wavelength ends, the cross spectra in Fig. 3(b) exhibit a wavelength variation that is an artifact of the normalization procedure as noted in Section 2. In particular, the increase in cross port signal above 1550 nm is simply a result of normalizing the relatively flat noise floor signal by the coupling envelope spectrum of the reference waveguide. The through port and cross port ratios D4/D1 and D4/D2 were also measured and found to be identical to the results in Fig. 3. The footprint of the crossings of approximately 100 μm × 100 μm is large compared with most other types of waveguide crossings. However, the device area can easily be reduced by reducing the free propagation length between collimators, and using collimators with shorter focal lengths.

## 5. Planar Waveguide Beam Splitter Couplers

Coupling a specific fraction of light from one single mode waveguide to another, or splitting the light into two output waveguides is an important function in integrated optical circuits. Y-junctions and multi-mode interference (MMI) couplers can be used to split light into two output waveguides but obtaining splitting ratios other than 50:50 is difficult. Both Y-junctions and MMIs can introduce scattering loss and back-reflection in high index contrast waveguide platforms, since there are always discontinuities along the optical transition from one to two waveguides [22,23]. Directional couplers are therefore more commonly used for coupling light from one waveguide into another in applications such as Si ring resonators and Mach-Zehnder interferometers [23-26]. Directional couplers consist of two waveguides separated by a small gap such that the light in one waveguide is evanescently coupled into the second waveguide. Since the waveguide separation can be varied adiabatically, directional couplers have very low reflection or scattering loss. The amount of light coupled to the second waveguide be set to any value from 50% to very small fractions of the incident power by adjusting the gap width and length of the coupled waveguide section. On the other hand, directional couplers are intrinsically wavelength dependent and the optimal coupling wavelength is sensitive to variations in the waveguide size, gap width and gap trench profile. When working with planar collimated beams, it is possible to use an alternative coupling that is the direct analog of a cube beam splitter used in bulk optics. The beam splitter coupler (BSC) structure can provide coupling with a very small wavelength dependence and no scattering or back reflection.

The BSC test structure is shown in Fig. 4, and is the same design as incorporated in the CGBR resonator of Ref. [15]. Here we extend that work to examine the effect of gap filling with SiO$_2$ cladding and also provide a comparison with theoretical models. The BSC coupler is a simple trench etched completely through the Si waveguide slab layer. The trench is oriented at 45° to the direction of a collimated incident beam. Since this angle is well within the total internal reflection regime, light cannot be scattered out of plane and is only reflected into the reflection port or evanescently coupled across the gap towards the transmission port. In this example the incident collimated beam is formed by parabolic collimator having a focal length F= 10 μm. The input waveguide tapers out to a 2 μm input waveguide aperture at the focus of the parabola. Identical collimators are used to focus the transmitted and reflected beams into the output waveguides. This configuration will produce a collimated beam width of 9 μm with a confocal parameter of $z_0$=114 μm. This is approximately the same as the distance from the input collimator to either the reflection and transmission collimators, so losses due to beam divergence are expected to be negligible. BSC couplers with gaps of 200 nm, 300 nm, 400nm and 500 nm were fabricated and tested. Assuming there is no power loss at the gap (e.g., from sidewall roughness scattering), the BSC transmission T can be taken as the ratio of transmitted power $P_1$ to total output power ($P_1+P_2$) from the two output ports, The fabricated gaps were chosen to target very small coupling ratios that would be used, for example, to build high quality factor ring resonators (i.e., T ≤ 1%). The splitting ratio will depend on the refractive index of the gap fill material. Here measurements were carried out both before the upper oxide cladding was deposited (an air-filled gap) and after deposition (an $SiO_2$ filled gap),

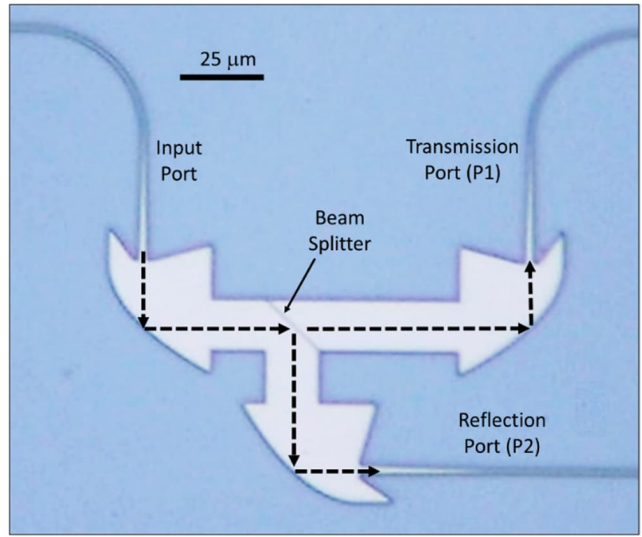

Fig. 4. An optical microscope view of the Si slab waveguide beam splitter coupler (BSC) test structure. The light areas are the Si waveguide regions. The dashed lines indicate the optical paths through the structure.

The transmission and reflection port spectra of each beam splitter were collected from λ=1480 nm to λ=1580 nm with TE polarized light, and normalized to the reference waveguide spectrum. Since the transmitted fraction is less than 2% of the input signal for all gaps, within the ±1 dB measurement uncertainty there was no measurable difference between reflection port spectra for the different splitter gaps. Similarly, for all gap widths the reflection port spectra were indistinguishable from the reference waveguide spectrum. This indicates that the insertion loss of the beam splitter structure is better than -1 dB, with no obvious wavelength dependence within the measured spectral range.

The coupling ratio spectra are shown in Fig. 5 for all devices. For an air-filled gap, the coupling ratios range from T = $7 \times 10^{-3}$ for the 200 nm gap to T = $8 \times 10^{-5}$ for the 500 nm gap. When the gap is filled with $SiO_2$, the coupling ratio increases in each case by approximately one order of magnitude, as would be expected given the longer evanescent field decay length for a higher gap index. Note that the air gap and filled gap spectra are for the same devices, measured before and after $SiO_2$ cladding deposition.

Fig. 6 presents a plot of the measured TE polarized transmission for each gap width at λ=1550 nm, both with and without $SiO_2$ gap fill. Also shown are model calculations using a simple effective index model, and using full three-dimensional FDTD simulations with a commercial FDTD tool (Lumerical). The effective index model treats the beam splitter as a bulk optic element comprising two blocks of material with the slab waveguide effective index $N_{eff} \sim 2.8$, separated by the air gap (index n=1) or oxide (index n=1.47) filled gap. Light is incident at 45°. The effective index model gives qualitative agreement with the measured dependence of transmission on the gap width and gap fill, but underestimates the transmission by a factor of two or more, with increasing discrepancies at larger gap widths. The effective index model is a purely two-dimensional model and cannot incorporate the details of the evanescent field distribution near the waveguide trench. The FDTD simulations again agree with the measured variations on gap width and gap fill material, but the FDTD simulation overestimates the transmission at each gap width. The differences between measurement and theory may arise from deviation in the fabricated gap widths from the nominal design values, although scanning electron microscope images indicate that the fabricated gap widths are within 25 nm of the nominal value. Another possible source of disagreement may arise from deviations of the sidewall of the fabricated trench edge from a perfectly vertical profile. Side wall angle, corner rounding, etc., are difficult to measure non-destructively and incorporate into a simulation model.

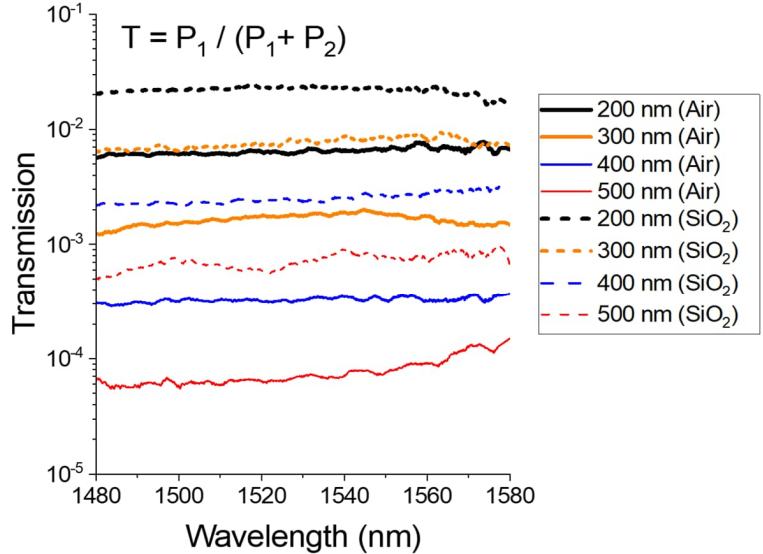

Fig. 5. Beam splitter transmission spectra for gap widths ranging from 200 to 500 nm, for both air-filled and SiO$_2$ filled gaps,

In principle, the BSC can be designed for any splitting ratio, but in practice splitting ratios with transmission larger than 10% require gap width of 100 nm or less. While this is within the scope of e-beam lithography and reactive ion etching processes, the relative accuracy of the etched gap width and side wall profile becomes difficult to control precisely below 100 nm widths. However, these beam splitters are well suited for attaining very low coupling ratios needed for high quality factor waveguide resonators where wider gaps are used. Beam splitters operate by evanescent coupling, and hence have a much weaker wavelength dependence than directional couplers, as evidenced by the data in Fig. 6. A theoretical estimate based on the effective index model suggests that the transmission varies by approximately 20% over a 100 nm wavelength range. This enables CGBR resonators to be designed to operate over a much wider wavelength range than traditional Si ring resonators.

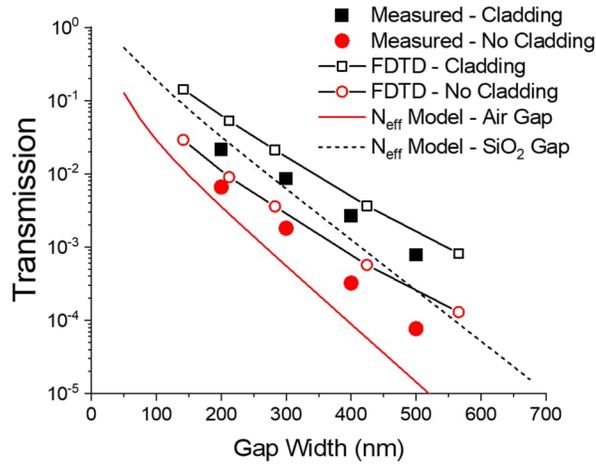

Fig. 6. The variation of the measured beam splitter coupler transmission with gap width at λ=1550nm, for air filled (no cladding) and SiO$_2$ filled (with cladding) gaps. Calculated results using a simple effective index model and a three-dimensional FDTD simulation are also shown.

Recently we have reported a high Q-factor planar slab waveguide circulating Gaussian beam resonator (CGBR) design that uses slab waveguide beam splitters as couplers [15]. Here we use a similar CGMR device to demonstrate confirm the wavelength independence of the BSC coupling ratio. The CGMR shown schematically in Fig. 7(a) has a cavity length of L= 100 μm, and uses two identical beam splitters with a 300 nm oxide filled gap as the add and drop couplers. The operating principles and details of the design are provided in Ref. [15], the only difference from the previous work being that this CGBR has an SiO$_2$ cladding and SiO$_2$ filled beam splitter gap. The drop port resonance spectrum is shown in Fig. 7(b), along with the measured FWHM (full width at half maximum) line widths across the 100 nm laser wavelength range. The resonator spectrum has been normalized by the reference waveguide spectrum to remove the wavelength dependence of the input and output grating couplers. The curvature of the system noise floor visible in Fig. 8(b) is again an artefact of this normalization procedure. The FWHM linewidth of an add-drop ring resonator can be calculated using the equation

$$FWHM = \frac{(1-r_a r_d \alpha)\lambda^2}{\pi N_g L \sqrt{r_a r_d \alpha}}. \qquad (3)$$

where $\alpha$ is the round-trip field attenuation and $r_a$ and $r_d$ are the through field coupling coefficients of the add and drop port couplers [25], and $N_g$ is the group effective index of the waveguide. These correspond to the field reflection coefficients in the beam splitter couplers shown in Fig. 5 and Fig. 7(a). The coupling coefficients can be expressed in terms of the beam splitter power transmission T such that $r_a = r_d = (1-T)^{1/2}$. The dashed curve shown in Fig. 7 (b) is the FWHM predicted by Eq. 3 assuming negligible round-trip loss ($\alpha =1$) and a transmission T=0.012. This transmission coefficient is close to but slightly larger than the T=0.009 value measured for the test structure with identical 300 nm oxide filled gap in Fig. 6. The difference in T between the resonator result and the direct beam splitter measurement may be due to the assumption that the cavity round trip loss is negligible. It is not possible to extract independent values for cavity loss and coupling from the data using Eq. 3, but an exact agreement between the directly measured structure in Fig. 6 and resonator derived splitting ratio from Eq. 3 is obtained by setting round trip field attenuation to $\alpha =0.997$ in Eq. 3. This corresponds to an effective propagation loss of 1.3 dB/cm in the cavity. This is comparable to typical measured losses for straight single mode Si waveguides. However, in the case of the CGBR resonator this effective propagation loss includes not only waveguide loss, but also the cumulative loss of 400 reflections (since each 100 μm cavity round trip involves four reflections), and any optical loss imposed by fundamental limitations the optical design. An example of the latter would be mode clipping of the tails of the Gaussian beam profile by the mirrors which by necessity have a finite aperture, and deviations of the circulating beam form a perfect Gaussian. The additional loss per mirror in the cavity (not including the out- coupling at the beam splitters) can be estimated to be $\alpha_{mirror} > 0.9985$, by assuming that the round-trip field loss of $\alpha =0.997$ is due to reflection loss alone. This confirms that all the side-wall mirrors are operating in near perfect TIR mode in the cavity and scattering losses are negligible.

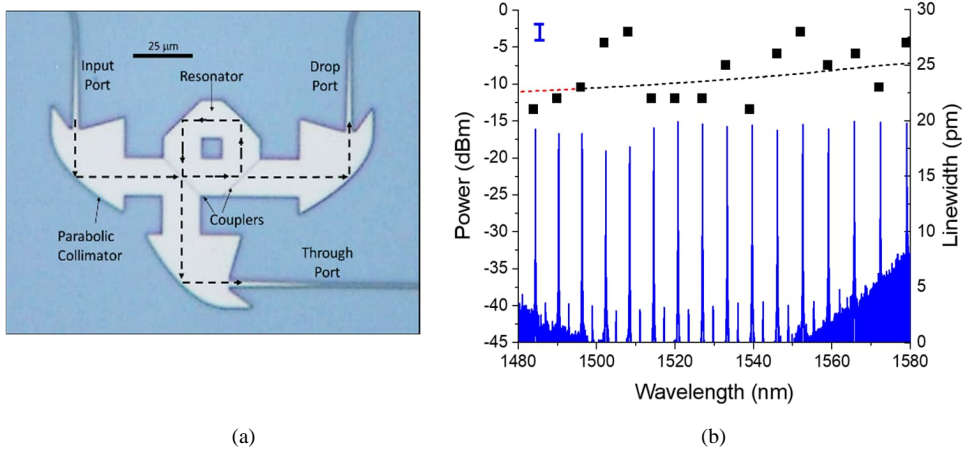

(a)          (b)

Fig. 7. (a) An optical microscope image of the circulating gaussian mode resonator layout for L= 100 μm cavity length and couplers with a 300 nm gap. The light areas are the Si slab waveguide regions. The dashed lines indicate the optical paths through the structure. (b) The drop port transmission spectrum (solid curve) and measured line widths (data points) for each resonance in the spectrum shown. The blue error bar indicates the peak power uncertainty due to ripple noise in the measurement system. The dashed curve indicates the calculated linewidths assuming a constant beam splitter coupling ratio.

The results shown in Fig. 7(b) show that the drop port spectrum resonance peak power and the resonance FWHM are approximately constant across the full 100 nm wavelength range. The weak positive wavelength slope of the FWHM theory is a result of the $\lambda^2$ term and group index dispersion in Eq. 3. The small random variations in FWHM and peak power from one resonance to the next from a background ripple noise in the measured spectra which has amplitude swings up to ±1 dB. This ripple is a result of scattering and back reflections in the measurement system and on-chip optical path. Since the ripple is random but with typical peak to peak wavelength separation comparable to the resonator linewidths, it cannot be easily removed from the data by a smoothing algorithm without altering the resonance line shapes. These results confirm that the beam splitter coupling ratio is wavelength independent across the 100 nm tuning range, and in this device a quality-factor of Q ~ 64000 can be maintained across this range and beyond.

## 6. Discussion

Several silicon waveguide components that operate without the use of single mode waveguides or lateral confinement have been described and characterized. These devices rely on mirrors formed by etching a shaped sidewall into the silicon waveguide layer. In previous work parabolic mirrors have been demonstrated as beam expanders [14] and as part of chip-optical fiber grating coupling structures [13]. In this work sidewall mirrors are used in a more general way to shape and direct collimated slab mode beams. Because of the high refractive index contrast of silicon waveguides, these mirrors operate in the TIR regime and are in principle lossless. It follows that similar mirror structures with different shapes can be used to carry out many different planar waveguide beam manipulations. The curved mirror used to form the Gaussian beam in the CGBR resonator cavity is one example. The only constraint is that incident beam angles are all larger than TIR critical angle. This constraint is relatively easy to satisfy in the silicon waveguide platform, since the critical angles are small: $\theta_c=21°$ for an unclad waveguide and $\theta_c=31°$ for a waveguide with oxide cladding.

One aspect TIR waveguide mirrors that has not been noted in previous work is that light undergoing TIR from an interface is reflected with an incident angle dependent phase shift $\delta$. In the case of TIR reflection of a plane wave at the interface between two bulk materials, this phase shift is given by the formula

$$\tan\frac{\delta}{2} = \frac{\sqrt{sin^2\theta - \left(\frac{n_2}{n_1}\right)^2}}{\left(\frac{n_2}{n_1}\right)^2 \cos\theta} \qquad (4)$$

where $n_2$ and $n_1$ are the refractive indices of the two materials [27]. Equation 4 applies for the case where the electric field is in the plane of incidence, as is the case for a TE polarized waveguide mode reflected from a sidewall mirror. For plane wave reflected from a planar interface, this phase shift is optically equivalent to a small displacement of the reflecting interface. For reflection of a diverging beam from a parabolic mirror, this is no longer the case. The angle of incidence changes across the beam and the phase front of the reflected beam will undergo a distortion. Applying a simple effective index-based calculation, the index $n_1$ can be set equal the Si waveguide effective index, and $n_2$ to the cladding index in Equation 2. This calculation predicts that for the incident angle range ($33° < \theta < 58°$) in the parabolic collimators used in this work, the TIR phase shift varies monotonically across the beam from $\delta = 0.82\pi$ to $\delta = 0.94\pi$ for a Si waveguide with an air cladding. In terms of wavelength, this corresponds to a phase front variation of $\Delta = 0.06\cdot\lambda$ across the collimated beam. For a Si waveguide with an $SiO_2$ cladding the predicted phase variation across the beam is four times larger at $\Delta = 0.23\cdot\lambda$. In the case of a finite width beam undergoing TIR, the reflected beam may also exhibit a lateral displacement from the expected specular beam path. This Goos-Haenschen shift has been the subject of extensive analysis and debate since it was first suggested by Isaac Newton [28,29]. For TIR reflection of a Gaussian beam with spot size $\omega_0$ and incident near the critical angle, Horowitz and Tamir [29] predicted that a reflected beam will undergo a lateral shift with an order of magnitude $d \sim (\omega_0\cdot \lambda)^{1/2}$. These phase effects are equivalent to a displacement and shape distortion of the reflecting interface, and could distort the reflected beam structure produced by the TIR mirrors. Eq. 4 and the work of Horowitz and Tamir [29] apply only to light reflected from a planar interface. The optical field distribution at a slab waveguide sidewall is considerably more complex than at a simple planar interface, and quantifying the TIR phase from a curved planar waveguide mirror will require extensive numerical modelling. But if they can be predicted with confidence, the TIR phase offsets should be easily corrected by adjusting the sidewall mirror shape to compensate for the phase shift.

In spite of the above considerations, there is no clear evidence that the TIR phase effects are responsible for measurable losses in the experimental results presented in this and previous work [13-15]. For structures involving short focal length (F < 20 μm) parabolas, the mirror losses in this work are below our measurement uncertainty (±1dB), consistent with the extremely low losses ( 0.15 dB) already reported by Xu et al. [13]. Nevertheless, the optical relay results in Section 3 show small increase in relay loss with increasing parabola size that is independent of the actual propagation distance. Whether this loss is related to the TIR phase effect or just simple imaging imperfections of the parabolic mirror-waveguide arrangement is not yet clear, but is an important subject for investigation.

In summary, the results presented here show that integrated optical functions can often be implemented using freely propagating beams shaped and manipulated by sidewall mirrors. The CGBR resonator results indicate that the reflection losses from straight sidewall mirrors forming the cavity are as low as -0.007 dB. This strategy is similar to building and optical layouts on a laboratory optical table, rather than building up conventional integrated optical circuits from single mode waveguides. By eliminating single mode waveguides and nanometer scale features, performance may be less sensitive to fabrication variations, losses and back-scattering due to sidewall roughness can be reduced, and sensitivity of optical phase to waveguide lateral dimension can be eliminated. Beam crossing can be essentially lossless with no back-reflection. Waveguide coupling using BSCs provides wavelength independent coupling over a 100 nm wavelength range. Our previous work on CGBR resonators [15] has already illustrated that this approach provides a path to achieving very high Q silicon waveguide resonators for sensors and wavelength filters. In general, a planar free propagating beam approach may also facilitate the integration of photonic chips with free space optics for new applications.

Accurate numerical simulations of planar slab waveguide devices with wide beams and long propagation distances can be time consuming. The results in this work show that simple working designs and qualitative performance prediction of the devices presented in this work can be accomplished using very simple Gaussian beam and effective index calculations. However, these simple models do not incorporate the details of the field distribution at the sidewall edge and the effect of the sidewall edge profile and angle, which will determine TIR phase shift and coupling across narrow beam splitter trenches. To assess the ultimate performance these devices can achieve, further work is required to develop the design rules and properties for the TIR mirror structures used to manipulate the beams, particularly the reflection phase, and also to go beyond the limitations of the simple Gaussian beam models which do not apply to more complex beam profiles.